\documentclass{rmf-d}
\usepackage{nopageno,rmfbib,multicol,times,epsf,amsmath,amssymb,cite}
\usepackage[latin1]{inputenc}
\usepackage[]{caption2}
\usepackage{graphics}
\usepackage{graphicx}
\usepackage[hidelinks]{hyperref}
\hypersetup{colorlinks, citecolor=black, linkcolor=black ,urlcolor={blue}}
\usepackage{wrapfig, blindtext}
\usepackage{comment}
\usepackage{tabularx}
\usepackage{float}
\usepackage{dblfnote}

\newcommand{\C}{{\kern+.25em\sf{C}\kern-.45em\sf{{\small{I}}} \kern+.45em\kern-.25em}}

\newcommand{\be}{\begin{equation}}
\newcommand{\ee}{\end{equation}}
\newcommand{\bea}{\begin{eqnarray}}
\newcommand{\eea}{\end{eqnarray}}
\newcommand{\nn}{\nonumber}

\newcommand{\la}{\langle}
\newcommand{\ra}{\rangle}
\newcommand{\ri}{{\rm i}}
\def\rmfcornisa{ \hfill\rmf{\bf }
\hfill }
\def\rmfcintilla{{\it Rev.\ Mex.\ Fis.\/} }
\clearpage \rmfcaptionstyle 
\begin{document}
\title{New insight in the 2-flavor Schwinger model based on lattice simulations
\vspace{-6pt}}
\author{Jaime Fabi\'{a}n Nieto Castellanos}
\author{Wolfgang Bietenholz}
\address{Instituto de Ciencias Nucleares, Universidad Nacional Aut\'{o}noma
de M\'{e}xico \\ A.P.\ 70-543, C.P.\ 04510 Ciudad de M\'{e}xico, Mexico}
\author{Ivan Hip}
\address{Faculty of Geotechnical Engineering, University of Zagreb \\
  Hallerova aleja 7, 42000 Vara\v{z}din, Croatia}
\maketitle
\vspace*{2mm}
\begin{abstract}
\vspace{1em}
We consider the Schwinger model with two degenerate, light fermion
flavors by means of lattice simulations. At finite temperature, we
probe the viability of a bosonization method by Hosotani {\it et al.}
Next we explore an analogue to the pion decay constant, which agrees
for independent formulations based on the  Gell-Mann--Oakes--Renner
relation, the 2-dimensional Witten--Veneziano formula and the
$\delta$-regime. Finally we confront several conjectures about the
chiral condensate with lattice results.

\vspace{1em}
\end{abstract}
\keys{Schwinger model, lattice gauge theory, finite temperature,
  chiral condensate, $\delta$-regime, pion decay constant \vspace{-4pt}}
\pacs{11.10.Kk, 11.10.Wx, 11.15.Ha, 12.20.-m \vspace{-4pt}}
\vspace*{1mm}
\begin{multicols}{2}

\section{The 2-flavor Schwinger model}

In the early 1960s, when quantum field theory was yet to be elaborated
as the correct theory of particle physics, Schwinger \cite{Schwinger}
analyzed Quantum Electrodynamics in $d=2$ space-time dimensions
(QED$_{2}$, or Schwinger model).
It shares qualitative properties with QCD, in particular
confinement, chiral symmetry breaking and topology.

Schwinger was particularly interested in the emergence of mass,
which was puzzling before the Higgs mechanism was established. In
fact, for $N_{\rm f}$ massless fermion flavors, the spectrum of QED$_{2}$
includes one massive and $N_{\rm f}-1$ independent massless bosons.
By analogy to QCD we denote them as the ``$\eta$-meson''
(which could also be interpreted as a massive ``photon'')
and the ``pions''. The $\eta$-mass was computed analytically
\cite{Belvedere},
\be  \label{metachiral}
m_{\eta}^{2} = \frac{N_{\rm f} g^{2}}{\pi} \ ,
\ee
where $g$ is the gauge coupling.

At a degenerate fermion mass $m>0$, there are conjectures but no
exact solutions for the masses $m_{\pi}$ and $m_{\eta}$. They can be
numerically measured with lattice simulations, which provide
fully non-perturbative results. We present such simulation results,
which we obtained with $N_{\rm f}=2$ degenerate flavors of dynamical
Wilson fermions, using the Hybrid Monte Carlo algorithm.
The renormalized fermion mass $m$ was measured based on the
PCAC relation.
Part of these results were anticipated in Ref.\ \cite{Lat21}.

\section{``Meson'' masses at finite temperature}

Bosonization reduces the Schwinger model to a quantum mechanical
system of $N_{\rm f}-1$ degrees of freedom; we call its temporal size
$L_{t}$. In the case of $N_{\rm f} =2$ degenerate flavors of mass $m$,
this method encodes the masses $m_\pi$ and $m_\eta$ in a Schr\"{o}dinger-type
equation for a periodic function $f(\varphi ) = f(\varphi + 2\pi)$
\cite{Hosotani},
\bea
\epsilon f(\varphi) &=&
\left( - \frac{d^2}{d\varphi^2} - \kappa\cos\varphi \right) f(\varphi) \ ,
\nn \\
\kappa &=& \frac{4}{\pi}m L_t \left[ B(m_{\eta} L_t ) B(m_\pi L_t) \right]^{1/2}
e^{-\pi/(2\mu L_t)} \ , \nn \\
B(z) &=& \frac{z}{4\pi}\textrm{exp}\left[ \gamma + \frac{\pi}{z}
  - 2\int_{1}^{\infty} \frac{du}{(e^{uz}-1)\sqrt{u^2 -1}}\right] \ , \nn \\
m_{\pi}^{2} &=& \frac{2 \pi^{2}}{L_{\rm t}^{2}} \kappa \int_{-\pi}^{\pi}
d\varphi \ \cos \varphi \ |f_{0}(\varphi)|^{2} \ , \nn \\
m_{\eta}^2 &=& m_{\pi}^{2} + \mu^{2} \ , \quad \mu = \sqrt{\frac{2}{\pi}} \, g \ ,
\label{eq:HosotaniEquations}
\eea
where $\gamma=0.577\dots$ is Euler's constant, $\epsilon$ is the energy,
and $f_0$ the ground state function.
This system of equations can be solved numerically \cite{Jaime},
but the viability of its solution is limited to $m \ll \mu$.

\begin{figure}[H]
\centering
\includegraphics[scale=0.5]{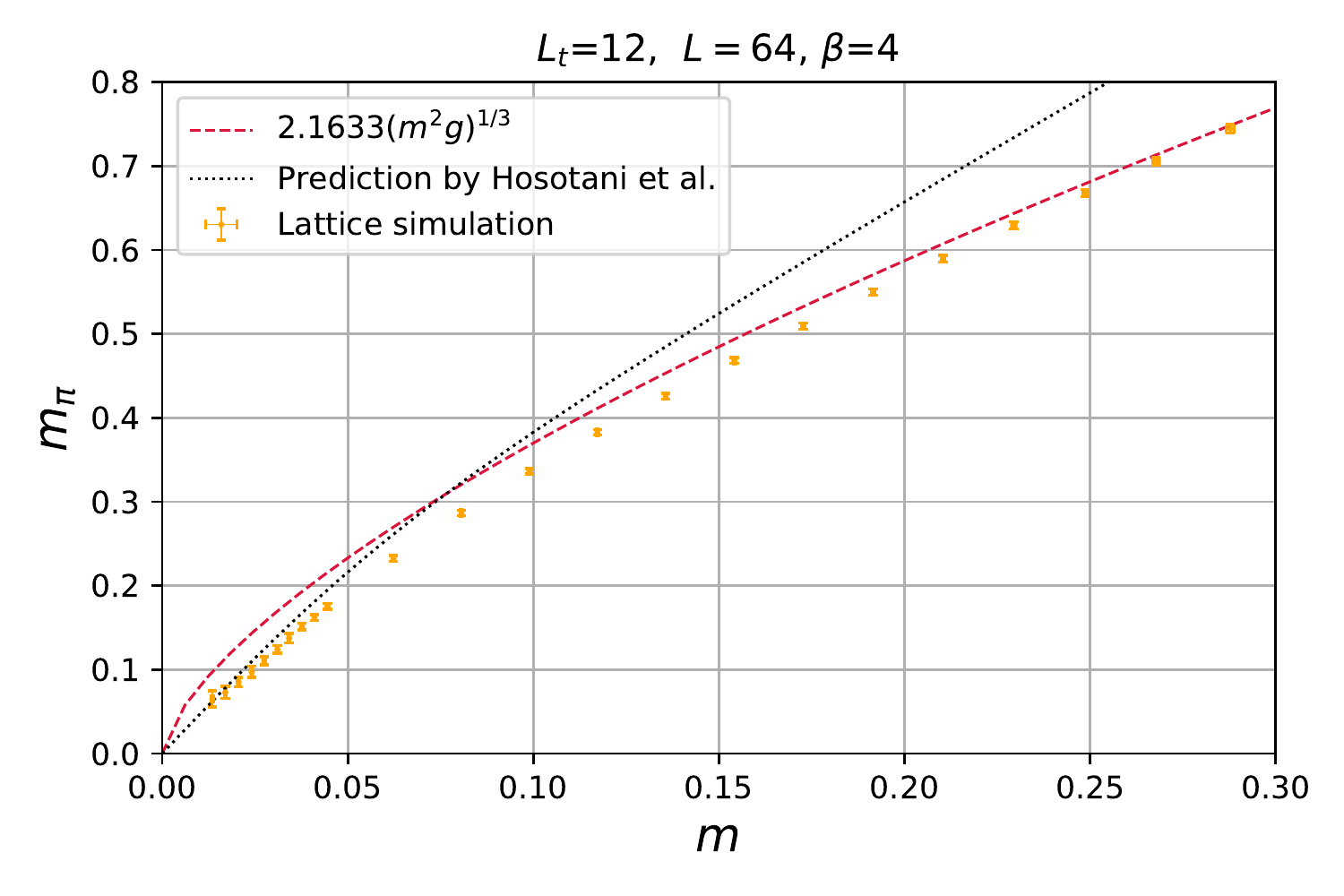}
\includegraphics[scale=0.5]{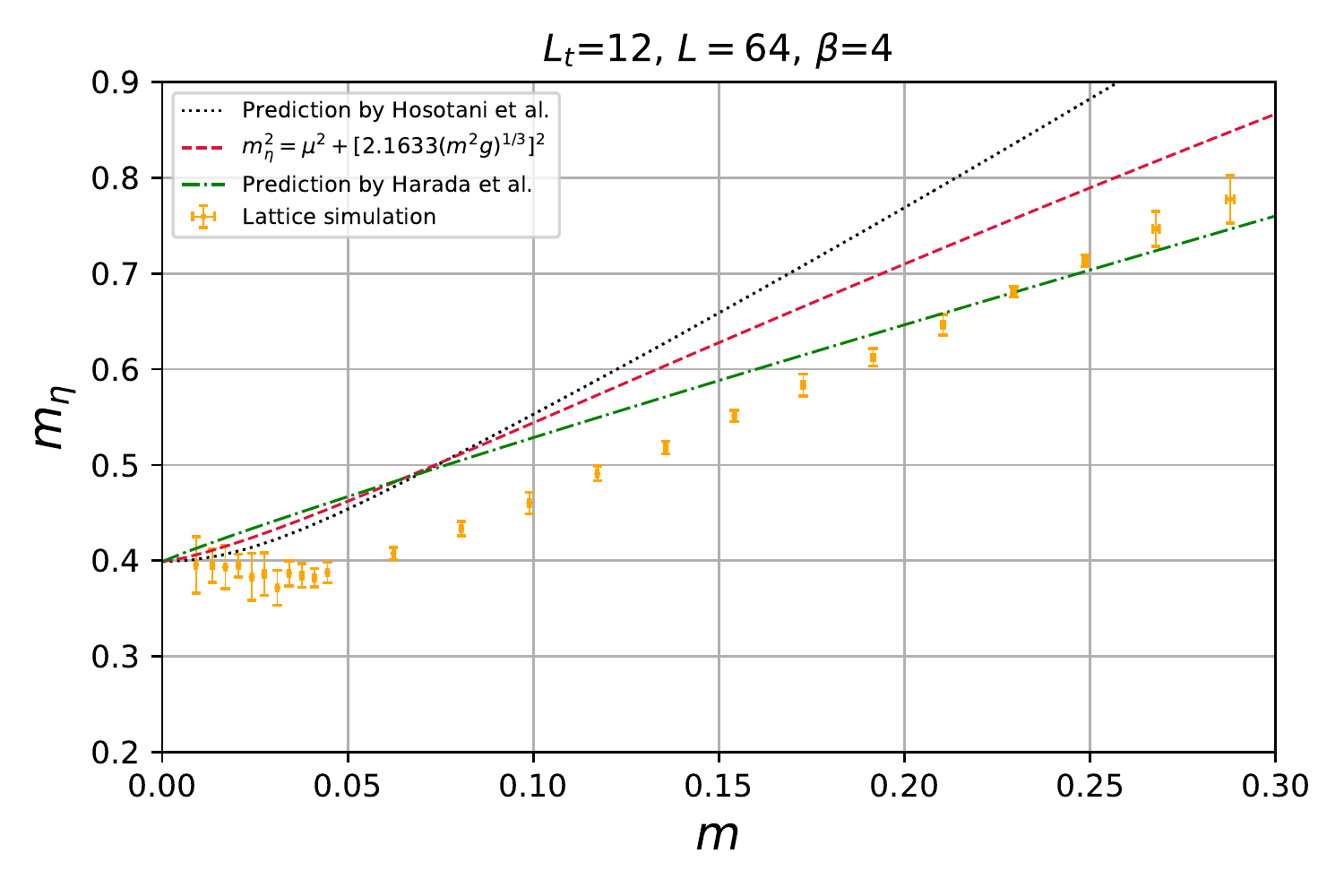}
\caption{The masses $m_\pi$ and $m_\eta$, depending on the
  renormalized fermion mass $m$.
  We show results obtained from Hosotani's equations
  \eqref{eq:HosotaniEquations}, from the asymptotic formula
  \eqref{asym}, from light-cone quantization \cite{Harada},
  and from lattice simulations.}
\label{fig:MpiMeta}
\end{figure}

In an infinite spatial volume, for a small mass $m$, the solution to
eqs.\ \eqref{eq:HosotaniEquations} for the ``pion'' mass takes the form
\be  \label{asym}
  m_\pi = 4 e^{2\gamma} \sqrt{\frac{2}{\pi}}(m^2 g)^{1/3}
  = 2.1633\dots (m^2 g)^{1/3}.
\ee
This is similar to  another infinite volume prediction by
Smilga \cite{Smilga97}, $m_\pi = 2.008\dots (m^2  g)^{1/3}$.
Figure \ref{fig:MpiMeta} compares the solution to Hosotani's
equations \eqref{eq:HosotaniEquations} and the asymptotic formula
\eqref{asym} to our simulation results on a lattice of size
$L_{t} \times L = 12 \times 64$, at $\beta \equiv 1/g^{2} =4$
(in lattice units), as a function of the fermion mass $m$.

We observe a quasi-chiral regime, with $m \lesssim 0.05$,
where the predictions for $m_{\pi}$ based on Hosotani's formula
is manifestly successful. There is another regime around
$0.25 \lesssim m \lesssim 0.3$ where the asymptotic formula
for $m_{\pi}$ agrees with the lattice data, but that could be
accidental, since the slopes differ.

On the other hand, the results for $m_\eta$ illustrate that none of the
predictions is accurate, except perhaps at tiny fermion mass, where
even the simple formula \eqref{metachiral} for the chiral limit
is more successful (it yields $m_{\eta} = 0.3989 \dots$).
In that case, we include a predictions, which is given
--- up to re-scaling --- in Ref.\ \cite{Harada},
$
m_{\eta} = \frac{g}{\sqrt{\pi}} [ e^{0.19} (\sqrt{\pi} m/g)^{0.993} + \sqrt{2}],
$
and which is (accidentally) close the the lattice data around $m \approx 0.23$.

\section{The ``pion decay constant''}

A frequent question about the multi-flavor Schwinger model refers
to the ``pions'' in the chiral limit, $m=0$: they are massless,
but in contrast to QCD they cannot represent Nambu-Goldstone bosons
due to the Mermin-Wagner-Coleman Theorem --- although at small $m>0$
they behave much like quasi-Nambu-Goldstone bosons. An explanation
is given {\it e.g.}\ in Ref.\ \cite{SmiVer}: at $m=m_{\pi}=0$ the
``pions'' do not interact.

Regarding the standard definition of the pion decay constant\footnote{Note
  that this ``pion'' does not actually decay.} $F_{\pi}$,
\be  \label{Fpistan}
\la 0 | J_{\mu}^{5}(0)|\pi (p) \ra = \ri p_{\mu} F_{\pi} \ ,
\ee
this property suggests $F_{\pi}(m=0)=0$.

However, there are other ways to define an analogue to $F_{\pi}$
in the 2-flavor Schwinger model, which lead to finite values.
We are going to see that they are quite consistent.

To the best of our knowledge, there is only one non-trivial prediction in
the literature for $F_{\pi}$ in the 2-flavor Schwinger model \cite{Harada}.
It is based on a light-cone quantization approach and it refers to the
relation
\be
\la 0 | \partial_{\mu} J_{\mu}^{5}(0)|\pi (p) \ra = m_{\pi}^{2} F_{\pi} \ ,
\ee
which we infer from eq.\ \eqref{Fpistan}, but this form allows for
$F_{\pi}(m=0)>0$. Harada {\it et al.} obtained a very mild dependence
on the (degenerate) fermion mass $m$ \cite{Harada},
\be  \label{FpiHarada}
F_{\pi} (m) = 0.394518(14) + 0.040(1) \, \frac{m}{g} \ .
\ee
Note that $F_{\pi}$ is dimensionless in 2 dimensions.

\subsection{Gell-Mann--Oakes--Renner relation}

The Schwinger model analogue of the Gell-Mann--Oakes--Renner relation
reads \cite{Smilga92}
\begin{equation}
F_{\pi}^{2}(m) = \frac{2 m \Sigma}{m_{\pi}^{2}} \ ,
\end{equation}
where $\Sigma = - \la \bar \psi \psi \ra$ is the chiral condensate.
Substituting infinite volume and low mass expressions
given in Ref.\ \cite{HHI95},
\begin{equation}
\Sigma = \frac{1}{\pi} \left( \frac{e^{4 \gamma} m \mu^2}{4} \right)^{1/3} \ ,
\quad m_\pi = (4e^{2\gamma} m^{2} \mu)^{1/3} \ ,
\end{equation}
yields
\be  \label{FpiHHI}
F_{\pi} = \frac{1}{\sqrt{2\pi}} = 0.3989 \dots \ .
\ee
This result is constant in $m$ and $g$, but for $m/g=0$ --- or close to
it --- we observe agreement with eq.\ \eqref{FpiHarada} up to $1\,\%$.
Ref.\ \cite{HHI95} also provides formulae for $\Sigma$ and $m_\pi$
in two other regimes, depending on $m$ and the volume. Inserting
either of them consistently leads again to eq.\ \eqref{FpiHHI}.
One might also insert the numerically measured values of $m_{\pi}$ and
$\Sigma$, see Section 4; this analysis is in progress.

\subsection{The 2d Witten--Veneziano formula}

In 't Hooft's formulation of large-$N_{\rm c}$ QCD, the Witten--Veneziano
formula \cite{WitVen} relates the $\eta'$-mass to the quenched
topological susceptibility $\chi_{\rm t}^{\rm q}$. In particular, it explains
--- as a topological effect --- why the $\eta'$-meson is so heavy
compared to the light meson octet. This relation involves the decay constant
$F_{\eta'}$, which --- at large $N_{\rm c}$ --- coincides with $F_{\pi}$.

According to Ref.\ \cite{GatSei}, the analogous relation in the
multi-flavor Schwinger model is actually more robust. 
In the chiral limit it reads
\be
m_{\eta}^{2} = \frac{2N_{\rm f}}{F_{\eta}^{2}} \chi_{\rm t}^{\rm q} \ .
\ee
If we employ relation \eqref{metachiral} along with the formula
\cite{SeiSta}
\be  \label{chitq}
\chi_{\rm t}^{\rm q} = \frac{g^2}{4\pi} \ ,
\ee
we obtain
\be
F_{\eta} = \frac{1}{\sqrt{2\pi}} \ .
\ee
If we further assume $F_{\eta} = F_{\pi}$, as in large-$N_{\rm c}$ QCD,
we arrive at exact agreement with eq.\ \eqref{FpiHHI}. 

Formula \eqref{chitq} is well tested as the continuum limit
of various lattice actions \cite{Lat21}, as requested by universality,
see also Refs.\ \cite{DurrHoelbling04}. On the other hand,
we are not aware of a sound justification for setting $F_{\eta} = F_{\pi}$
in the Schwinger model.

\subsection{The $\delta$-regime}

We proceed to yet another, independent way of introducing an analogue
to $F_{\pi}$. Here we refer to the $\delta$-regime, which was introduced
in QCD by Leutwyler \cite{Leutwyler}. It is characterized by a small
spatial volume, but a large extent in Euclidean time,
\be
V = L^{3} \times L_{\rm t} \ ,
\quad L \lesssim \frac{1}{m_\pi} \ll L_{\rm t} \ .
\ee
As a finite-size effect, there is a residual pion mass $m_{\pi}^{\rm R}$
even in the chiral limit,
\be
m_{\pi}^{\rm R} = m_{\pi}(m =0) > 0 \ .
\ee
It was computed to leading order in Ref.\ \cite{Leutwyler}, and
to next-to-leading order --- in a general space-time dimension
$d \geq 3$ --- in Ref.\ \cite{Hasenfratz93},
\bea
m_{\pi}^{\rm R} \!\!&=& \!\! \frac{N_\pi}{2\Theta} \ , \nn \\
\Theta  \!\!&=&\!\! F_{\pi}^{2} L^{d-1} \left[1 +
    \frac{N_\pi-1}{2\pi F_{\pi}^{2} L^{d-2}}
    \left( \frac{d-1}{d-2} + \dots \right) \right] \ . \qquad
\label{mpiRHas}
\eea
$N_{\pi}$ is the number of pions (or generally of Nambu-Goldstone
bosons), and if we consider the system as quasi-1d quantum mechanics,
$\Theta$ represents a moment of inertia.

Formula \eqref{mpiRHas} is not designed for $d=2$, where there are
no Nambu-Goldstone bosons and the next-to-leading term is singular.
We try to apply it nevertheless, restricting the formula to
Leutwyler's leading order, and interpreting $N_{\pi}$ as the
number of ``pions'' in the $N_{\rm f}$-flavor Schwinger model,
$N_{\pi} = N_{\rm f} -1$. Thus we conjecture for $N_{\rm f}=2$
\be  \label{eqmpiR}
m_{\pi}^{\rm R} \simeq \frac{1}{2F^2_{\pi} L} \ .
\ee
If the behavior $m_{\pi}^{\rm R} \propto 1/L$ is confirmed, the proportionality
constant provides another way of introducing $F_{\pi}$, by means of
yet another analogy to QCD.

\begin{figure}[H]
\vspace*{-3mm}
\centering
\includegraphics[scale=0.5]{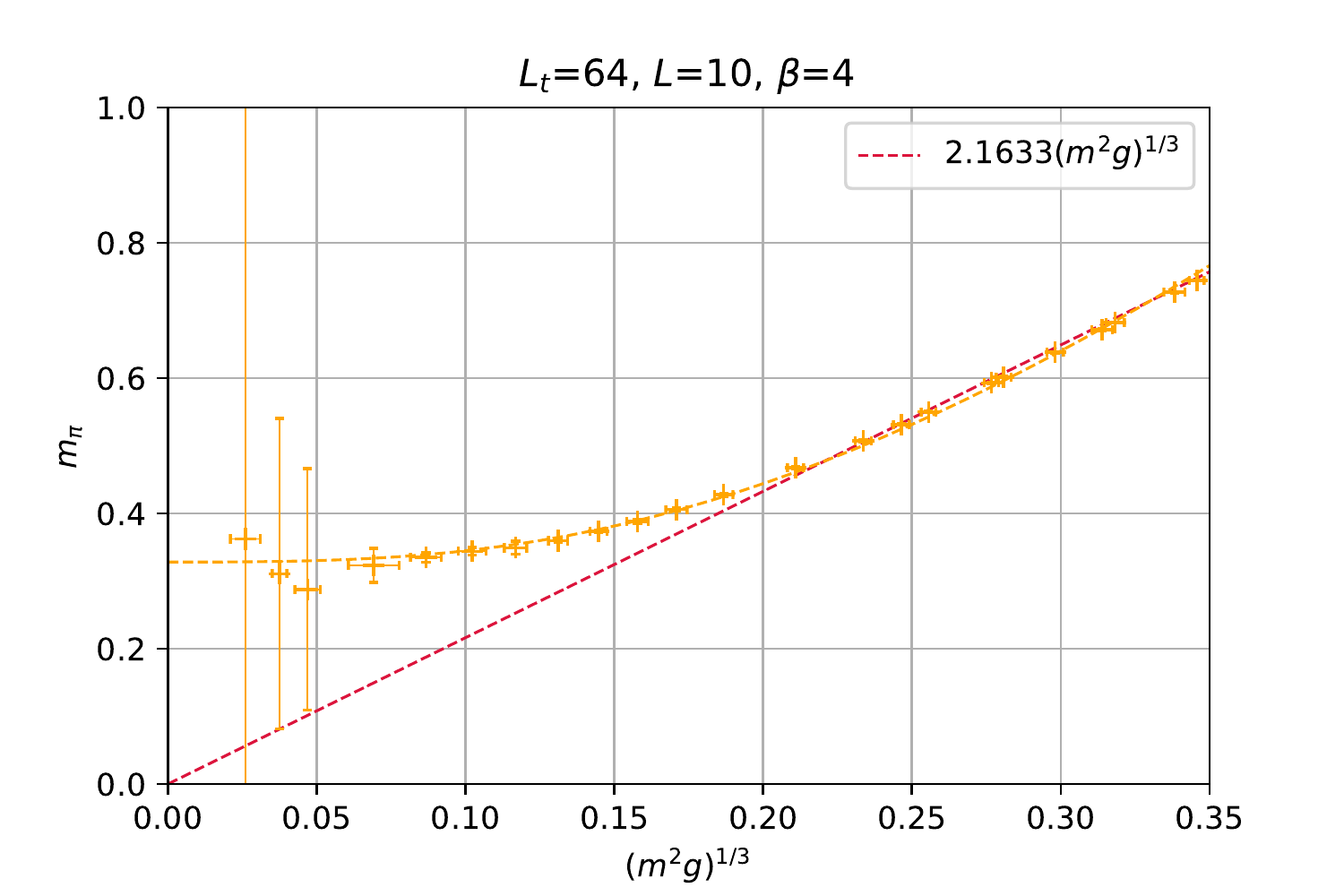}
\caption{Pion mass as a function of $(m^2 g)^{1/3}$,
  in a volume $L \times L_{\rm t} =10 \times 64$,
  at $\beta =4$. As expected, the error bars shoot up at tiny $m$, but
  the data at moderate $m$ guide a controlled chiral extrapolation
to $m_{\pi}^{\rm R}$.}
\label{fig:mpiR}
\vspace*{-3mm}
\end{figure}

In order to probe this scenario, we measured $m_{\pi}$ at different
values of $m$, and extrapolated to the chiral limit in order to obtain
simulation results $m_{\pi}^{\rm R}$ (simulations directly at tiny
$m$ are plagued by notorious technical problems).
Figure \ref{fig:mpiR} shows an example for such an extrapolation.

\begin{figure}[H]
\vspace*{-3mm}
\centering
\includegraphics[scale=0.46]{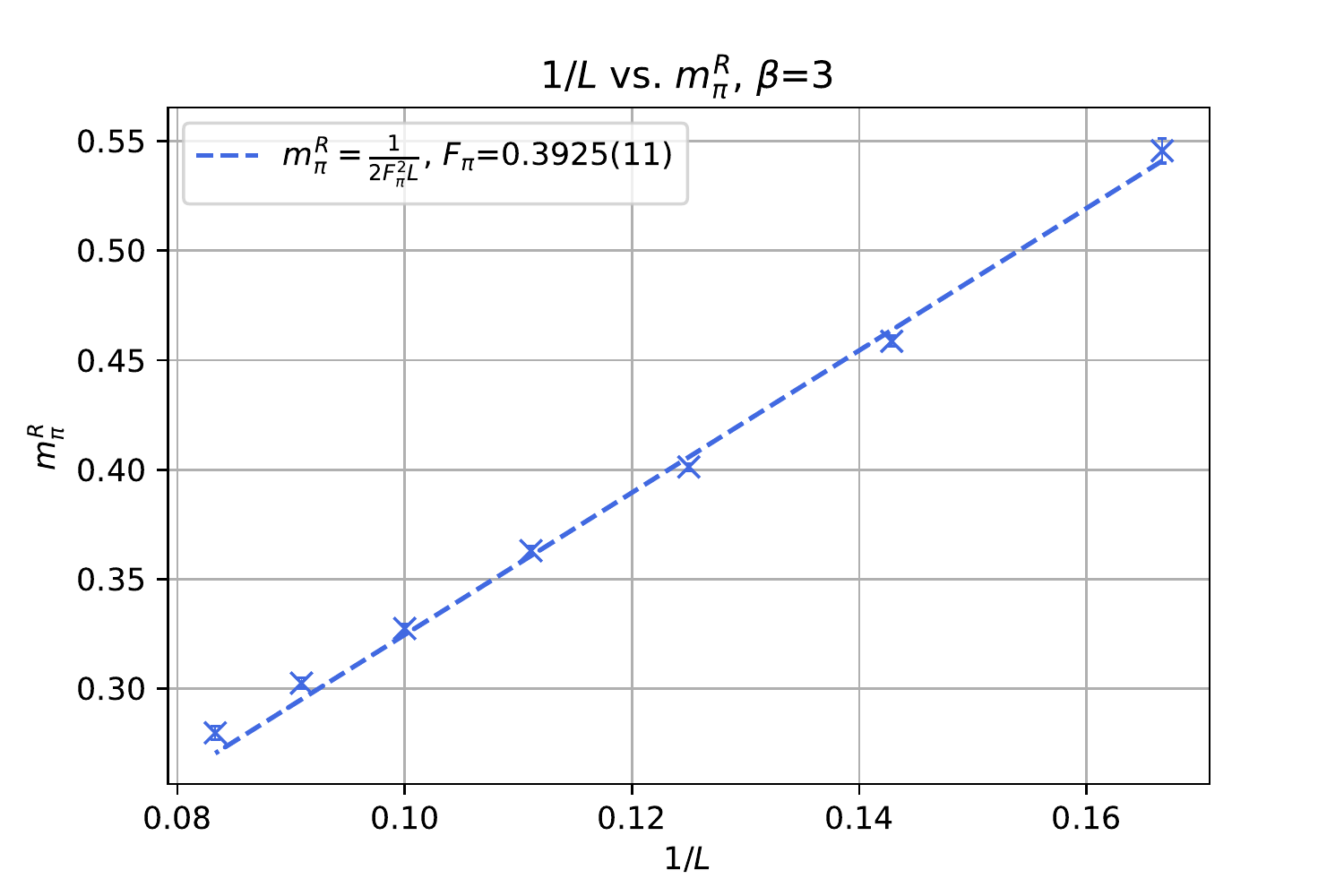}
\includegraphics[scale=0.46]{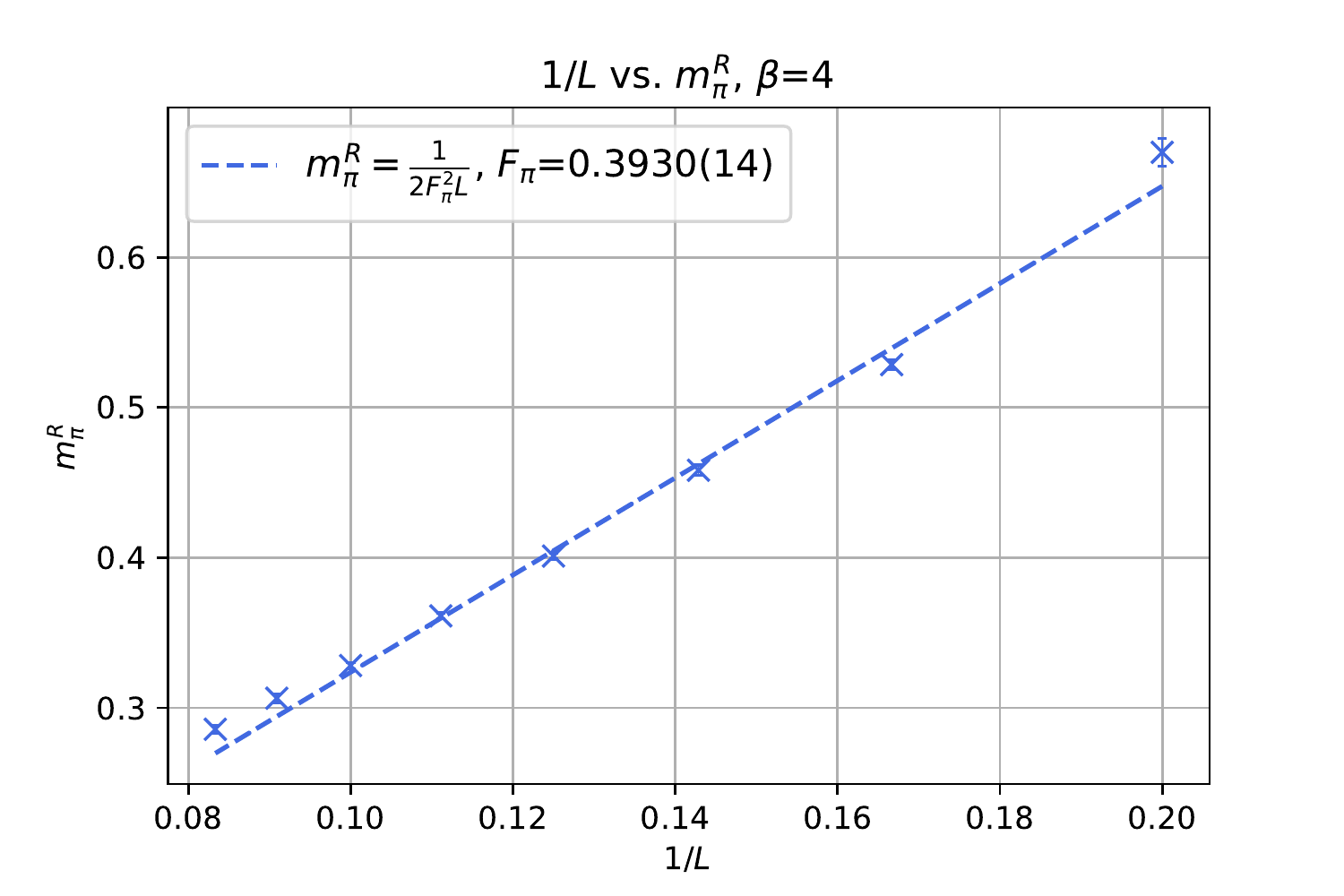}
\includegraphics[scale=0.46]{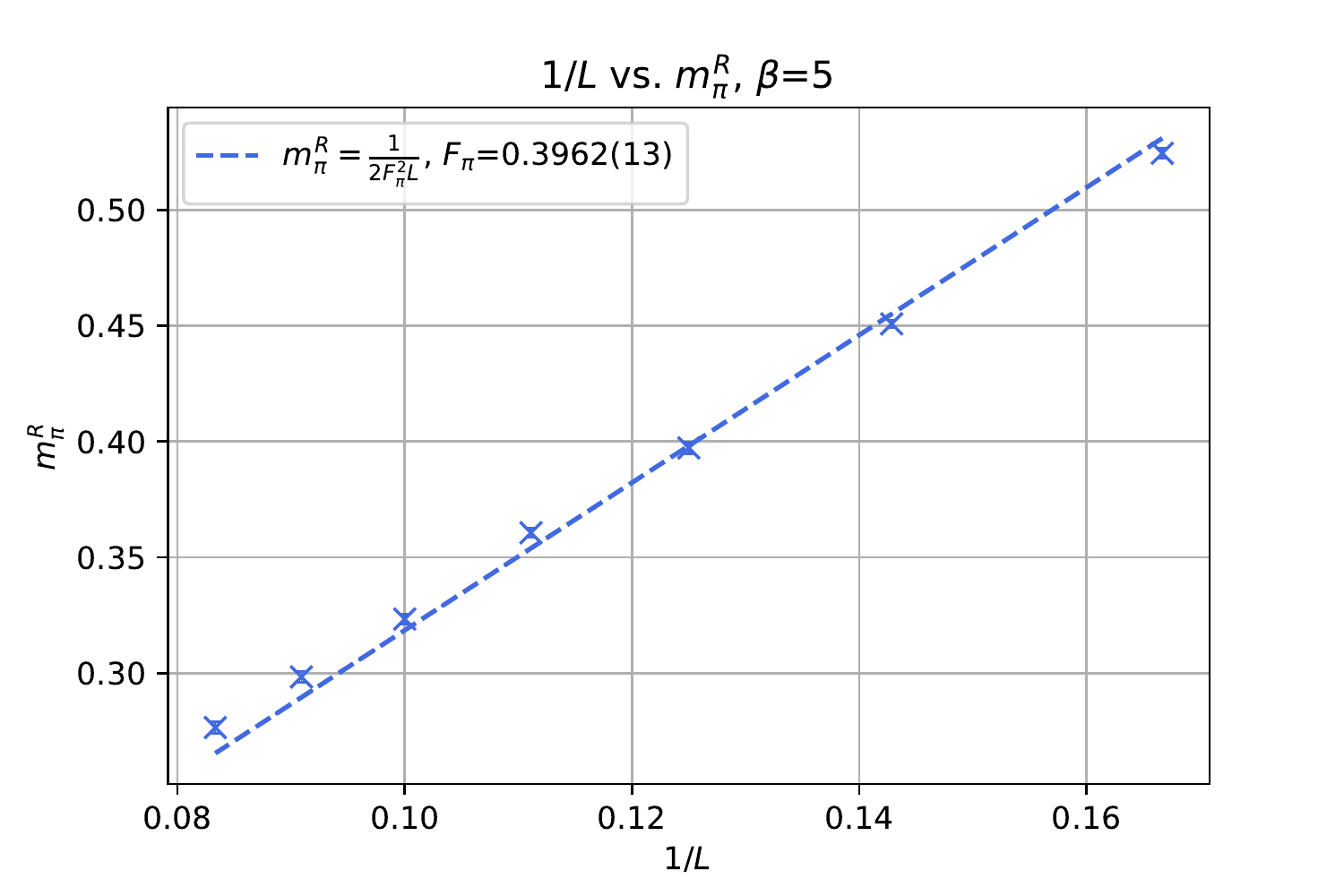}
\caption{Residual pion mass as function of $1/L$ for
  $\beta =3$, $4$ and $5$. We observe good agreement with the
conjectured relation $m_{\pi}^{\rm R} \propto 1/L$.}
\label{fig:mpiRL}
\end{figure}

Repeating this extrapolation at different spatial sizes $L$ leads to
good agreement with the conjectured proportionality relation
$m_{\pi}^{\rm R} \propto 1/L$, as Figure \ref{fig:mpiRL} shows for
three $\beta$-values.

This allows us to employ eq.\ \eqref{eqmpiR} and extract the value of
$F_{\pi}$ --- as defined in the $\delta$-regime. The results at
$\beta = 3$, $4$ and $5$ are given in Table \ref{tab:FpiResults}.
\begin{table}[H]
\centering
\begin{tabular}{|c|c|}
\hline
$\beta$ & $F_\pi$  \\
\hline
3       & 0.3925(11)  \\
4       & 0.3930(14)  \\
5       & 0.3962(13)  \\
\hline
\end{tabular}
\vspace*{2mm}
\caption{$F_\pi$ obtained from $m_{\pi}^{\rm R}$ in the $\delta$-regime,
  for three different $\beta$-values.}
\label{tab:FpiResults}
\end{table}
They agree very well for various values of $\beta$. In addition, they
are very close to the result for $F_{\pi}$ of Ref.\ \cite{Harada}, and
to the results that we obtained based on the 2d Gell-Mann--Oakes--Renner
relation, and on the Witten--Veneziano formula
(if we assume $F_{\pi} = F_{\eta}$).

\section{The chiral condensate}

If we rely on the result \eqref{FpiHHI}, we can re-write the
Gell-Mann--Oakes--Renner relation in the form
\be  \label{SigmaGMOR}
\Sigma = \frac{m_\pi^2}{4\pi m},
\ee
and use it to extract a value for the chiral condensate $\Sigma$
from the numerical solution of eqs.\ \eqref{eq:HosotaniEquations}.
\begin{figure}[H]
\centering
\includegraphics[scale=0.6]{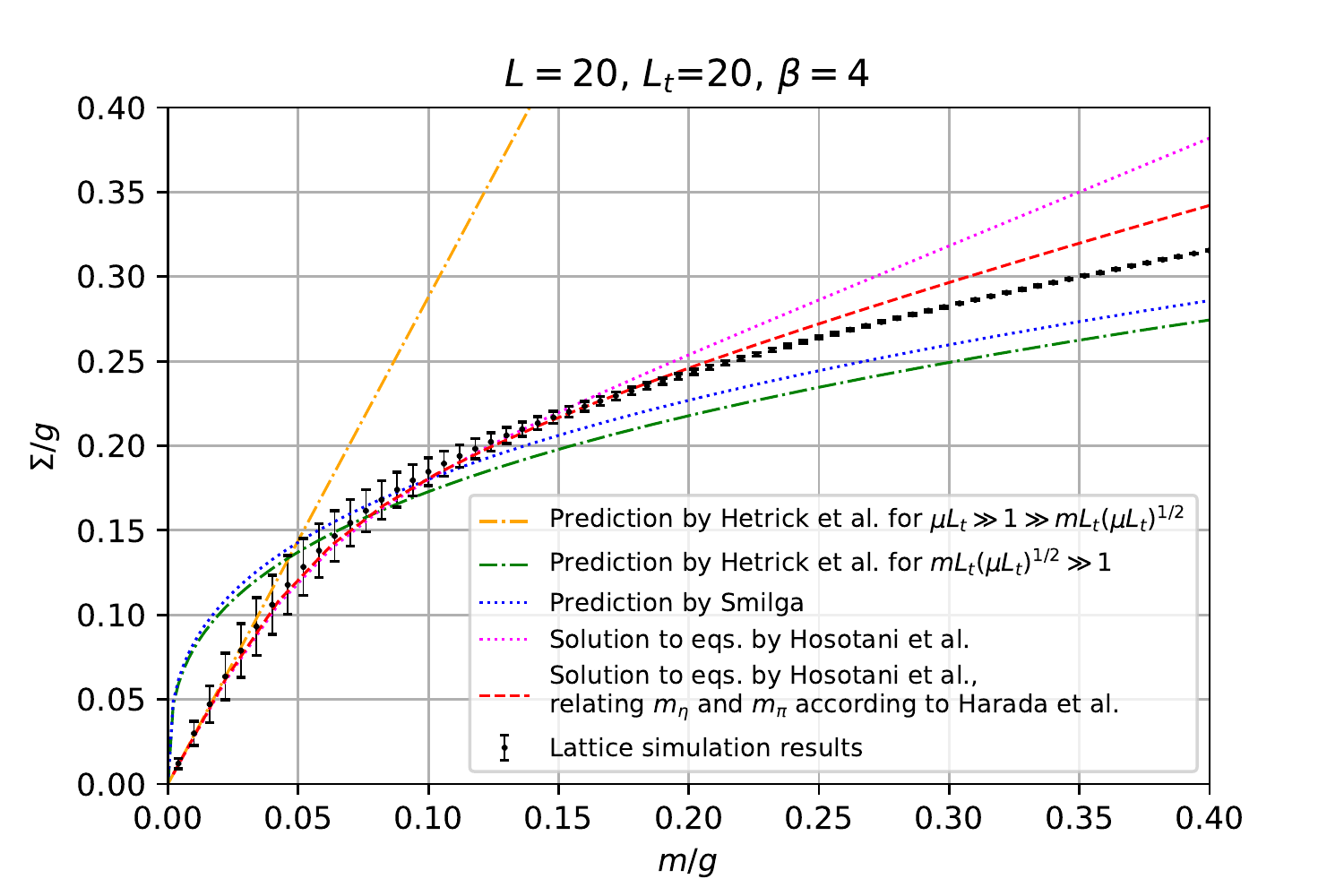}
\caption{A multitude of results for the chiral condensate $\Sigma$.}
\label{fig:PbP}
\end{figure}

Figure \ref{fig:PbP} compares results for $\Sigma$ by a
variety of approaches: the line linear in $m/g$ is the tangent
at $m/g=0$, which was correctly predicted in Ref.\ \cite{HHI95}.
For our parameters, the corresponding conditions
mean $8 \gg 1$ (which is plausible), and $m/g \ll 0.035$.
However, the data follow well this straight line even up to
$m/g \approx 0.035$.

We add another small-mass predictions given in Ref.\ \cite{HHI95},
which refers to the regime $m/g \gg 0.035$, hence it should be
valid in most of this plot, but it does not agree well with the
lattice data (which are obtained with overlap fermions).

Next we show the line for a similar prediction from Ref.\ \cite{Smilga97},
which is not in accurate agreement with the lattice results either.

The solution of eqs.\ \eqref{eq:HosotaniEquations}, inserted
in relation \eqref{SigmaGMOR}, works better. It can be further
improved if we replace the
(somewhat troublesome) ansatz $m_{\eta}^{2} = m_{\pi}^{2} + \mu^{2}$
by a formula for $m_{\eta}$,
which is derived in Ref.\ \cite{Harada}.

Unlike Figs.\ \ref{fig:MpiMeta} to \ref{fig:mpiRL}, here we obtain
an almost continuous line of lattice results, because we are using
a single set of quenched configurations which are re-weighted by the
fermion determinant to compute the chiral condensate for different
fermion masses. As the ratio $m/g$ grows,
we see that none of the theoretical predictions is truly successful,
as we already observed in the case of the ``meson'' masses.
We should add, however, that all these formulae refer to the continuum
and an infinite spatial size, so the discrepancies could be (in part)
due to finite-size effects and lattice artifacts.

\section{Summary and mysteries}

We presented lattice simulation results for the ``meson'' masses
and the chiral condensate in the 2-flavor Schwinger model, and
confronted them with a multitude of theoretical predictions.
While some of them work well at small fermion mass $m$, none of
them is truly successful at moderate $m$.

We also formulated the pion decay constant $F_{\pi}$ in various ways, 
by referring to different analogies to QCD. For three formulations
we obtained consistent values, which are compatible with
$F_{\pi}(m=0) = 1/\sqrt{2\pi}$, and with a previous study in
Ref.\ \cite{Harada}. This is very satisfactory, but there remain
two open questions: 1. Why does the method based on the Witten--Veneziano
formula require the identification $F_{\pi} = F_{\eta}\,$? 2. How can this
be reconciled with relation \eqref{Fpistan}, which suggests
$F_{\pi}(m=0) =0\,$?\\

\noindent
{\bf Acknowledgments:} We thank the organizers of the
{\it XXXV Reuni\'{o}n Anual de la Divisi\'{o}n
de Part\'{\i}culas y Campos} of the {\it Sociedad Mexicana
de F\'{\i}sica}, where this talk was presented by JFNC.
We further thank Stephan D\"{u}rr and Christian Hoelbling
for useful communication.
The code development and testing were performed at the cluster Isabella
of the Zagreb University Computing Centre (SRCE). The production runs
were carried out on the cluster of the Instituto de Ciencias Nucleares,
UNAM.
This work was supported by UNAM-DGAPA through PAPIIT project IG100219,
``Exploraci\'{o}n te\'{o}rica y experimental del diagrama de fase de
la cromodin\'{a}mica cu\'{a}ntica'', by the
Consejo Nacional de Ciencia y Tecnolog\'{\i}a (CONACYT),
and by the Faculty of Geotechnical
Engineering (University of Zagreb, Croatia) through the project ``Change
of the Eigenvalue Distribution at the Temperature Transition''
(2186-73-13-19-11).

\end{multicols}
\medline
\begin{multicols}{2}

\end{multicols}
\end{document}